\documentclass[12pt]{article}

\begin{document}

\title{A new way to deal with Izergin-Korepin determinant at root of unity}
\author{Yu.~G.~Stroganov\\
\small \it Institute for High Energy Physics\\[-.5em]
\small \it 142284 Protvino, Moscow region, Russia}
\date{}

\maketitle

\begin{abstract}
I consider the partition function of the inhomogeneous 6-vertex model defined 
on the $n$ by $n$ square lattice.
This function depends on 2n spectral parameters $x_i$ and $y_i$ attached to the
horizontal and vertical lines respectively. 
In the case of domain wall boundary conditions it is given by Izergin-Korepin
determinant. 
For $q$ being a root of unity the partition function satisfies to a special linear
functional equation.
This equation is particularly good when the crossing parameter $\eta=2\pi/3$.
In this case it can be used for solving some of the problems related to the 
enumeration of alternating sign matrices. 
In particular, it is possible to reproduce the refined ASM distribution discovered
by Mills, Robbins and Rumsey and proved by Zeilberger. 
Further, it is well known that the partition function is symmetric in the $\{x\}$
and as well in the $\{y\}$ variables. 
I have found that in the case of $\eta=2\pi/3$, the partition function is symmetric
in the union $\{x\} \cup \{y\}$!
This nice symmetry is used to find some relations between the numbers of such
alternating sign matrices of order $n$ whose two '1' are located in fixed positions
on the boundary of the matrices.
Finally I derive the equation giving `top-bottom double refined' ASM distribution. 
\end{abstract}

\newpage 

\begin{center}{\bf1. Six-vertex model with domain wall boundary and Izergin-Korepin
determinant}
\end{center}

Let us consider the inhomogeneous six-vertex model on a square lattice.
The states of the model are given by assigning arrows to each edge of an 
$n \times n$ square lattice so that at each vertex, two arrows go in and two go out
(\it ice condition \rm).
Spectral parameters $\{x_1,x_2,...,x_{n}\}$ and $\{y_1,y_2,...,y_{n}\}$ are attached
to the horizontal and vertical lines respectively.
Edges point inward at the sides and outward at the top and bottom as in figure 1.
With these boundary conditions which are called \it domain wall  boundary conditions
\rm \cite{kor1} the states of the model are in in one-to-one correspondence with 
a certain set of matrices, called \it alternating sign matrices \rm
.

\vspace{0.5cm}

\begin{picture}(0,200)
\put(40,60){$x_n$}
\put(40,150){$x_2$}
\put(40,180){$x_1$}

\put(60,60){\vector(1,0){18}}
\put(78,60){\line(1,0){12}}

\put(60,150){\vector(1,0){18}}
\put(78,150){\line(1,0){12}}

\put(60,180){\vector(1,0){18}}
\put(78,180){\line(1,0){12}}

\put(90,60){\line(1,0){60}}
\put(90,150){\line(1,0){60}}
\put(90,180){\line(1,0){60}}

\put(160,60){\dots}
\put(160,150){\dots}
\put(160,180){\dots}

\put(180,60){\line(1,0){30}}
\put(180,150){\line(1,0){30}}
\put(180,180){\line(1,0){30}}

\put(240,60){\vector(-1,0){18}}
\put(222,60){\line(-1,0){12}}

\put(240,150){\vector(-1,0){18}}
\put(222,150){\line(-1,0){12}}

\put(240,180){\vector(-1,0){18}}
\put(222,180){\line(-1,0){12}}

\put(88,20){$y_1$}
\put(118,20){$y_2$}
\put(208,20){$y_{n}$}

\put(90,60){\vector(0,-1){18}}
\put(90,42){\line(0,-1){12}}

\put(120,60){\vector(0,-1){18}}
\put(120,42){\line(0,-1){12}}

\put(210,60){\vector(0,-1){18}}
\put(210,42){\line(0,-1){12}}

\put(90,60){\line(0,1){30}}
\put(120,60){\line(0,1){30}}
\put(210,60){\line(0,1){30}}

\put(89,100){\vdots}
\put(119,100){\vdots}
\put(209,100){\vdots}
%\put(90,100){\vdots}
%\put(120,100){\vdots}
%\put(210,100){\vdots}

\put(90,120){\line(0,1){60}}
\put(120,120){\line(0,1){60}}
\put(210,120){\line(0,1){60}}

\put(90,180){\vector(0,1){18}}
\put(90,198){\line(0,1){12}}

\put(120,180){\vector(0,1){18}}
\put(120,198){\line(0,1){12}}

\put(210,180){\vector(0,1){18}}
\put(210,198){\line(0,1){12}}

\put(160,100){$\ddots$}
\put(120,0){Figure 1} 

\end{picture}

\vspace{0.5cm}

There are six possible configurations of the arrows on the edges for a given vertex as
in figure 2. 

\vspace{0.5cm}

\begin{picture}(0,60)
\put(0,40){\vector(1,0){13}}
\put(13,40){\line(1,0){7}}
\put(20,40){\vector(1,0){13}}
\put(33,40){\line(1,0){7}}
\put(20,20){\vector(0,1){13}}
\put(20,33){\line(0,1){7}}
\put(20,40){\vector(0,1){13}}
\put(20,53){\line(0,1){7}}
\put(43,38){\tiny =}
\put(70,40){\vector(-1,0){13}}
\put(57,40){\line(-1,0){7}}
\put(90,40){\vector(-1,0){13}}
\put(77,40){\line(-1,0){7}}
\put(70,60){\vector(0,-1){13}}
\put(70,47){\line(0,-1){7}}
\put(70,40){\vector(0,-1){13}}
\put(70,27){\line(0,-1){7}}
\put(93,38){\tiny = \small a}
\put(120,40){\vector(1,0){13}}
\put(133,40){\line(1,0){7}}
\put(140,40){\vector(1,0){13}}
\put(153,40){\line(1,0){7}}
\put(140,40){\vector(0,-1){13}}
\put(140,27){\line(0,-1){7}}
\put(140,60){\vector(0,-1){13}}
\put(140,47){\line(0,-1){7}}
\put(163,38){\tiny =}
\put(190,40){\vector(-1,0){13}}
\put(177,40){\line(-1,0){7}}
\put(210,40){\vector(-1,0){13}}
\put(197,40){\line(-1,0){7}}
\put(190,40){\vector(0,1){13}}
\put(190,53){\line(0,1){7}}
\put(190,20){\vector(0,1){13}}
\put(190,33){\line(0,1){7}}
\put(213,38){\tiny = \small b}
\put(240,40){\vector(1,0){13}}
\put(253,40){\line(1,0){7}}
\put(280,40){\vector(-1,0){13}}
\put(267,40){\line(-1,0){7}}
\put(260,40){\vector(0,-1){13}}
\put(260,27){\line(0,-1){7}}
\put(260,40){\vector(0,1){13}}
\put(260,53){\line(0,1){7}}
\put(283,38){\tiny =}
\put(310,40){\vector(-1,0){13}}
\put(297,40){\line(-1,0){7}}
\put(310,40){\vector(1,0){13}}
\put(323,40){\line(1,0){7}}
\put(310,60){\vector(0,-1){13}}
\put(310,47){\line(0,-1){7}}
\put(310,20){\vector(0,1){13}}
\put(310,33){\line(0,1){7}}
\put(333,38){\tiny = \small c}
\put(160,5){Figure 2}
\end{picture}

\vspace{0.5cm}
  
The Boltzmann weights which are assigned to every vertex of the lattice have the form:
\begin{eqnarray}
\label{eq1}
&&a(x-y)=\frac{\sin (x-y+\eta/2)}{\sin \eta}, \quad  b(x-y)=\frac{\sin
(x-y-\eta/2)}{\sin
\eta},\\
&&c(x-y)=1. \nonumber 
\end{eqnarray}
The different letters in Figure 1 correspond to the different functions introduced
in equation (\ref{eq1}). The $x$ and $y$ are the horizontal and vertical spectral
parameters, respectively, which depend on the vertex position. 
The $\eta$, sometimes called \it crossing parameter \rm has the same value for all
vertices.

The weight of a state of the model is the product of the weights of its vertices,
and the \it partition function (state sum) \rm is the total weight of all states.

Let $Z_n(x_1,x_2,...,x_{n}; y_1,y_2,...,y_{n})$ be the resulting state sum.  
Izergin \cite{iz1} (using the above mentioned work of Korepin 
\cite{kor1} ) found a determinant representation for Z.
\begin{eqnarray}
\label{eq2}
&&Z_n(\{x\},\{y\})=\prod_{1 \le i,j \le {n}} \sin(x_i-y_j+\eta/2) \>
\prod_{1 \le i,j \le {n}} \sin(x_i-y_j-\eta/2) \times \nonumber \\
&& \times \frac{ \mbox{det} M(\{x\},\{y\})} 
{\prod_{1 \le i < i^{\prime} \le {n}} \sin(x_i-x_{i^{\prime}})
\prod_{1 \le j < j^{\prime} \le {n}} \sin(y_j-y_{j^{\prime}})} 
\end{eqnarray}
where the entries of an $n \times n$  matrix $M(\{x\},\{y\})$ are
\begin{eqnarray}
\label{eq3}
&&M_{i,j}=\frac{1}{\sin(x_i-y_j+\eta/2) \> \sin(x_i-y_j-\eta/2)},\quad i,j=1,2,...,n.
\end{eqnarray}

Let us list some properties of the state sum which are used in this paper (see, for
example,~\cite{KIB} or~\cite{{Ku1}}).

The function $Z_n(\{x\},\{y\})$:
\begin{itemize}
\item
 is symmetric in the $\{x\}$ and in the $\{y\}$. 
\item
 is a trigonometric polynomial 
of degree at most $n-1$ in $u$, with $\pi$ being
quasiperiod:
\begin{eqnarray}
&&Z(u+\pi) = (-1)^{n-1} Z(u). \nonumber
\end{eqnarray}
where $u$ is one of the variables $x_1,x_2,...,x_{n}; y_1,y_2,...,y_{n}$.
\item
 is left unchanged by transformation 
\begin{eqnarray}
&& x \rightarrow \pi/2-x,\quad y \rightarrow -y,\quad \eta \rightarrow \pi-\eta, \nonumber
\end{eqnarray}
 while function $b$ (see equations (\ref{eq1})) changes its sign. 
 \end{itemize}

\vspace{0.5cm}

\begin{center}{\bf2. The basic equation for the state sum}
\end{center}

Let us fix the crossing parameter  
\begin{eqnarray}
&&\eta=\frac{2\pi}{N},    \nonumber 
\end{eqnarray}
where $N$ is an integer, $N \ge 3$.

Consider the set of the matrices $M^{(k)}$ given by equation (\ref{eq3}) with the 
 spectral parameter $x_1$ shifted by $k\eta$:
\begin{eqnarray}
\label{eq4}
&&M^{(k)} \equiv M(x_1+k\eta,x_2,...x_n;y_1,y_2,...y_n), \quad k=0,...,N-1.    
\end{eqnarray}
The variable $x_1$ occurs only in the upper row of the matrices $M^{(k)}$:
\begin{eqnarray}
\label{eq5}
&&M^{(k)}_{1,j}=\frac{1}{\sin(x_1-y_j+ (k+1/2)\eta) \> \sin(x_1-y_j+(k-1/2)\eta)}, \\
&&\quad j=1,2,...,n.\nonumber
\end{eqnarray}
Consider the sums of entries of the upper row over the set of $M^{(k)}$ 
\begin{eqnarray}
\label{eq6}
&&\sum_{k=0}^{N-1} M^{(k)}_{1,j}, \quad j=1,2,...,n.
\end{eqnarray}
Inserting (\ref{eq5}) into (\ref{eq6}) and using a  simple identity
\begin{eqnarray}
&&\frac{\sin \eta} {\sin (u+\eta /2)\>\sin(u-\eta /2)} = \frac{\exp i(u-\eta /2)}{\sin
(u-\eta /2)} -
\frac{\exp i(u+\eta /2)}{\sin (u+\eta /2)}, \nonumber
\end{eqnarray}
we obtain that all sums (\ref{eq6}) are equal to 0.
Since $\mbox {det} M^{(k)}$  for every $j$ linearly depends on $M^{(k)}_{1,j}$ this implies
that 
\begin{eqnarray}
\label{eq7}
&&\sum_{k=0}^{N-1} \mbox {det} M^{(k)} 
\equiv \sum_{k=0}^{N-1} \mbox {det} M(x_1+k\eta,x_2,...x_n;y_1,y_2,...y_n)=0.    
\end{eqnarray}
Using representation
(\ref{eq2}) for the state sum,
we immediately obtain the basic equation

\begin{eqnarray}
&&\sum_{k=0}^{N-1} \biggl\{ Z(x_1+k\eta,x_2,...x_n;y_1,y_2,...y_n) \prod_{i=2}^n
\sin(x_1-x_i+k\>\eta)
\nonumber \\
&& \prod_{j=1}^{n} \biggl [ \sin(x_1-y_j+(k+1/2) \eta ) \> \sin(x_1-y_j+(k-1/2)
\eta)\biggr
]^{-1}\biggr\}=0.
\nonumber
\end{eqnarray}
This formula appears to be useful for the study of the partition function $Z$.

In this paper we limit ourselves to the case 
\begin{eqnarray}
&&N=3, \quad \eta=2\pi/3, \nonumber
\end{eqnarray}
Using an identity
\begin{eqnarray}
&&\sin(u+\frac{\pi}{3}) \> \sin(u-\frac{\pi}{3}) \>\sin u = -\frac{1}{4}\sin 3u,
\nonumber 
\end{eqnarray}
we obtain a much more handy equation
\begin{eqnarray}
\label{eq12}
&& \sum_{k=0}^{2}\biggl \{ Z(x_1+2\pi k/3,x_2,...x_n;y_1,y_2,...y_n) \nonumber \\
&&\prod_{i=2}^n \sin(x_1-x_i+2\pi k/3) 
 \prod_{j=1}^{n} \sin(x_1-y_j+2\pi k/3)\biggr \}=0.
\end{eqnarray}
Let us write
\begin{eqnarray}
\label{xy_u}
&& u\>\> \mbox{for}\>\> x_1,\nonumber \\
&& u_i\>\> \mbox{for}\>\>  x_{i+1}, \quad i=1,...,n-1, \\
&&\mbox{and} \>\>u_{i+n-1}\>\>\mbox{for}\>\> y_i, \quad  i=1,...,n. \nonumber
\end{eqnarray}

After introducing the function $f$ defined by
\begin{eqnarray}
 \label{eq14}
&&f(u) = Z(u)\> \prod_{i=1}^{2n-1} \sin (u-u_i),
\end{eqnarray}
we can rewrite equation (\ref{eq12}) in the following form:
\begin{eqnarray}
 \label{eq15}
&&f(u)+f(u+\frac{2\pi}{3})+f(u+\frac{4\pi}{3})=0,
\end{eqnarray}
where, for brevity, we write $f(u)$ and $Z(u)$ rather than $f(u,u_1,...u_{2n-1})$
and $Z(u,u_1,...u_{2n-1})$ suppressing the variables $u_i,\>\>i=1,...,2n-1$ which remain
unchanged throughout. 

\bf The point is that equations (\ref{eq14}) and (\ref{eq15}) 
can be used to find  the state sum $Z(u)$ up to an arbitrary constant.
\rm

Using the properties of the state sum listed in Section 1 we obtain
that the function $f(u)$  is a trigonometric polynomial of degree at most $3n-2$ which 
satisfies:
\begin{eqnarray}
\label{addref}
&&f(u+\pi) = (-1)^{n} f(u).
\end{eqnarray}
This polynomial is expressed  by a finite Fourier sum:
\begin{eqnarray}
 \label{eq18}
&&f(u) = \sum_{k=1}^{3n-1} b_k\>\exp\{i(3n-2k)u\}.
\end{eqnarray}
where coefficients $b_k,\quad k=1,2,...,3n-1$ depend on the variables $u_i,\>i=1,...,2n-1$.
Inserting this expression  into (\ref{eq15}) 
we find that every third coefficient is zero:
\begin{eqnarray}
 \label{eq19}
&&b_{3\kappa} = 0,\quad  \kappa=1,2,....n-1.
\end{eqnarray}

Let $u_i \ne u_j$ for $i \ne j$.
In this case the remaining $2n$ coefficients can be found up to normalization by solving a
homogeneous system of linear equations\footnote{This problem is a special case
of interpolation problem.}:
\begin{eqnarray}
\label{hom}
&& f(u_j)=\sum_{k=1,k \ne 3\kappa}^{3n-1} \exp\{i(3n-2k)u_j\}\>\>b_k\>=0, \quad
j=1,2,...2n-1,
\end{eqnarray}
since the rank of the coefficient matrix is $2n-1$ for generic values of $u_j$. 

The resulting $f(u)$ is given (up to an arbitrary constant) by the determinant
\begin{eqnarray}
&&P= \mbox{det} \>\>\begin{array}{|ccccc|}
t^{3n-2}&t_1^{3n-2}&t_2^{3n-2}&\dots&t_{2n-1}^{3n-2} \\
t^{3n-4}&t_1^{3n-4}&t_2^{3n-4}&\dots&t_{2n-1}^{3n-4}  \\
t^{3n-8}&t_1^{3n-8}&t_2^{3n-8}&\dots&t_{2n-1}^{3n-8}  \\
\dots&\dots&\dots&\dots&\dots   \\
t^{2-3n}&t_1^{2-3n}&t_2^{2-3n}&\dots&t_{2n-1}^{2-3n}
\end{array}  \nonumber
\end{eqnarray}
where $t=\exp (i\>u)$ and $t_j=\exp (i\>u_j),\>j=1,2,...,2n-1$.
Indeed, let us evaluate the determinant by the left column expansion:
\begin{eqnarray}
&&P(u) = \sum_{k=1,k \ne 3\kappa}^{3n-1} \exp\{i(3n-2k)u\}\>\>{\tilde b}_k\>,\nonumber
\end{eqnarray}
where ${\tilde b}_k$ are the corresponding cofactors depending on the variables $u_j,\quad
j=1,2,...,2n-1$. 
We see first that the determinant $P(u)$ is expressed by a finite Fourier sum of the same
kind as (\ref{eq18}).
Second, condition (\ref{eq19}) is satisfied. 
At last it is obvious that $P(u=u_j)=0,\>j=1,2...,2n-1$ and 
hence the coefficients ${\tilde b}_k$ satisfy the system of equations (\ref{hom}). 
Therefore the ratio
\begin{eqnarray}
 \label{crat}
&&P(u)/f(u)= C(u_1,u_2,...,u_{2n-1})
\end{eqnarray}
does not depend on the variable $u$.

\begin{center}{\bf3. The symmetry of the state sum in the spectral parameters}
\end{center}

Inserting definition (\ref{eq14}) into equation (\ref{crat}) we obtain
\begin{eqnarray}
 \label{eq22}
&&P(u;u_1,u_2,...,u_{2n-1})=\tilde{C}(u_1,u_2,...,u_{2n-1})\nonumber \\
&&  \prod_{j=1}^{2n-1} \sin (u-u_j) \prod_{1\le j <
j^{\prime} \le 2n-1}
\sin (u_j-u_j^{\prime})\>\>Z(u;u_1,u_2,...,u_{2n-1}). 
\end{eqnarray}
Picking out the factor $\prod_{1\le j < j^{\prime} \le 2n-1} \sin
(u_j-u_j^{\prime})$ we emphasize the
symmetry of the functions entering equation (\ref{eq22}). 
Let us return to the parameters $\{x\}$ and $\{y\}$ (see definition 
(\ref{xy_u})). Both the left hand
side of equation (\ref{eq22}) and the product on the right hand side are  antisymmetric in
the
union
$\{x\}\cup
\{y\}$.  Combining
this property with the above-mentioned symmetry of the state sum Z(\{x\},\{y\}) in the
$\{x\}$
and in the $\{y\}$
separately, we obtain that the unknown function $\tilde{C}$ has to be symmetric  
in the $\{x\}$ and in the $\{y\}$ as well.
From the other hand this function does not depend on the variable $u \equiv x_1$. 
Consequently it does not depend on any of the the variables $x_i,\>\>i=1,2,...,n$. 

Returning to the beginning of Section 2 and using the left column
instead of the upper row, we obtain that  the function $\tilde{C}$ does not depend on
variables $y_i,\>\>i=1,2,...,n$ as well. Hence $\tilde{C}$ is a constant which possibly
depends on $n$.
So we get the nice result that

\bf{for $\eta=2\pi/3$, the partition function $Z(\{x\},\{y\})$ is symmetric in the union
$\{x\}\cup \{y\}$.} \rm

\begin{center}{\bf4. The refined ASM conjecture}
\end{center}

 Each of the further sections address some aspects of the alternating sign
 matrices (ASMs) of size n, which are in one-to-one correspondence with the
 states of the model we consider \cite{MRR},\cite{EKLP}.
The number of ASMs, for example, is  equal to the state sum $Z$ with $a=b=c=1$.

Consider the case when all spectral parameters, except one are equal to zero:
\begin{eqnarray}
\label{1case}
&& x_1=u ,\nonumber \\
&& x_i=0,\quad i=2,3,...,n,\\
&& y_j=0,\quad j=1,2,...,n. \nonumber
\end{eqnarray}
In the upper row of the lattice the Boltzmann weights are given by:
 \begin{eqnarray}
\label{upperBW}
&&a(u)=\frac{2}{\sqrt{3}} \sin (\pi/3+u), \quad b(u)=\frac{2}{\sqrt{3}} \sin
(\pi/3-u),\\
 &&c(u)=1 \nonumber.
\end{eqnarray}

All the Boltzmann weights in the remaining rows are equal to 1.

Let us divide the set of the states into $n$ classes according to the column number `r' of 
a (unique) vertex of the upper row where the horizontal arrows change their direction.
Let $A(n,r)$ be the number of the states of the $r^{th}$ class, then the partition
function (the state sum) is given by:
\begin{eqnarray}
\label{1stZ}
&&Z(u)=\sum_{r=1}^n A(n,r)\> a^{r-1}(u)\> b^{n-r}(u).
\end{eqnarray}

In terms of ASMs, it is also the number of ASMs  of order $n$ whose sole '1' of the upper
row, is at the $r^{th}$ column.
The wonderful story of these numbers can be found in the book by Bressoud~\cite{Br}.
Mills, Robbins and Rumsey discovered  a nice formula for
$A(n,r)$~\cite{MRR2,MRR}.
Their conjecture which is known as the refined ASM conjecture was proven by
Zeilberger~\cite{Z}, who found the state sum $Z(u)$ extending Kuperberg's
method~\cite{Ku1}. 
They both began with Izergin-Korepin determinant and solved a difficult problem, since
both the determinant and the product in the denominator of equation (\ref{eq2}) `badly'
vanish.

We also intend to find $A(n,r)$ using equations (\ref{eq14}), (\ref{eq15})
and general properties of the state sum $Z(u)$ listed in Section 1.  
It does not take a lot of time, because the corresponding problem has been  already
solved  in paper~\cite{S}.

Let us return to equation (\ref{eq19}). We have established that the function
$f(u)$ can be presented  by the finite Fourier sum (\ref{eq18}). Taking into account
that some of the coefficients $b_k$ are equal zero (see equation (\ref{eq19})) we can
rewrite it to the form
\begin{eqnarray}
&& f(u)=\sum_{m=0}^{n-1} (b_m^+e^{i(4-3n+6m)u}+b_m^-e^{-i(4-3n+6m)u}),\nonumber
\end{eqnarray}
In contract with the previous case we cannot use equations (\ref{hom}).
Instead we use the fact that in the last case (see equation (\ref{1case})) the monomial
$\sin^{2n-1}u$
divides this function due to definition (\ref{eq14}):
\begin{eqnarray}
\label{simplef}
&&f(u)\equiv Z(u) \sin^{2n-1}u.
\end{eqnarray}
This condition fixes the coefficients $b^{\pm}_m, m=0,1,...,n-1$ up to an arbitrary constant.
Indeed, the first $2n-2$ derivatives of the function $f(u)$ have to be zero at $u=0$:
\begin{eqnarray}
&&f^{(l)}(u)|_{u=0}=\sum_{m=0}^{n-1} (b_m^++(-1)^l b_m^-)(4-3n+6m)^{l}=0,\nonumber \\
&&l=0,1,...,2n-2.
\nonumber
\end{eqnarray}
If $l$ is even then we obtain the system
\begin{eqnarray}
&&\sum_{m=0}^{n-1} (b_m^++ b_m^-)(4-3n+6m)^{2\lambda}=0, \quad \lambda=0,1,...,n-1, \nonumber
\end{eqnarray}
which (from the non-singularity of the Vandermond matrix) implies
\begin{eqnarray}
&&b_m^++ b_m^-=0, \quad m=0,1,...,n-1. \nonumber
\end{eqnarray}
The odd derivatives give
\begin{eqnarray}
&&\sum_{m=0}^{n-1} b^+_{m} (4-3n+6m)^{2\lambda +1}=0,\quad
\lambda=0,1,...,n-2. \nonumber
\end{eqnarray}
This system is equivalent to the condition that the relation 
\begin{eqnarray}
\label{eq30}
&&\sum_{m=0}^{n-1}b^+_{m} (4-3n+6m)\> p((4-3n+6m)^2)=0
\end{eqnarray}
is valid for all polynomials $p(x)$ of degree $n-2$.
Consider the $n-1$ special polynomial of degree $n-2$:
 \begin{eqnarray}
&&p_{\mu}(x)=\prod_{m=1,m \ne \mu}^{n-1} (x-(4-3n+6m)^2),\quad
\mu=1,2,...,n-1.
\nonumber
\end{eqnarray}
Inserting these polynomials into (\ref{eq30}) we find simple relations connecting
$b^+_m$  with $b^+_0$. It is possible to write the answer in terms of binomial coefficients,
namely,
\begin{eqnarray}
\label{eq32}
&&f(u)=\mbox{const}\>\sum_{m =0}^{n-1}
\biggl(\begin{array}{c}
n - \frac{4}{3} \\
m
\end{array}\biggr)
\biggl(\begin{array}{c}
n - \frac{2}{3} \\
n - m -1
\end{array}\biggr)
\sin(4-3n+6m)u.
\end{eqnarray}
 This function satisfies the second order linear differential equation
 \begin{eqnarray}
\label{eq33}
&&f^{\prime \prime} - 6(n-1)\cot 3u\> f^{\prime} -(3n-2)(3n-4)f=0.
\end{eqnarray}
We return now to equation (\ref{1stZ}). Combining it with equation (\ref{simplef})
we obtain
\begin{eqnarray}
\label{fnew}
&&f(u)=\sin ^{2n-1}u \>\sin^{n-1}(u+\pi/3) \> \sum_{r=1}^n A(n,r)\biggl
(\frac{b(u)}{a(u)}\biggr )^{n-r}.  
\end{eqnarray}
Inserting the last equation into ODE (\ref{eq33}) and changing variable to 
 \begin{eqnarray}
&&t=\frac{b(u)}{a(u)}=\frac{\sin(\pi/3-u)}{\sin(\pi/3+u)}, \nonumber 
\end{eqnarray}
we obtain the second order linear differential equation 
 \begin{eqnarray}
\label{eq37}
&&t(1-t) A^{\prime \prime} +2(1-n-t) A^{\prime} +n(n-1) A=0,
\end{eqnarray}
for the generating function:
\begin{eqnarray}
\label{eq38}
&&A(t)=\sum_{r=1}^n A(n,r)\>t^{n-r}.
\end{eqnarray}
Inserting this sum into differential equation (\ref{eq37})  we get a recursion relation:
\begin{eqnarray}
&&(2n-r-1)\>r\>A(n,r+1) = (n-r)\>(n+r-1)\>A(n,r), \nonumber
\end{eqnarray}
which is satisfied by
\begin{eqnarray}
\label{eq40}
&&A(n,r)=\frac{A(n,1)}{(2n-2)!}\>\frac{(n+r-2)!\>(2n-r-1)!}{(r-1)!\>(n-r)!}.
\end{eqnarray}

It is clear that the number of ASMs of order $n$ with `1' at the left upper corner
$A(n,1)$ is equal to
the total number of ASMs of order $n-1$:
\begin{eqnarray}
&&A(n,1)=A_{n-1}. \nonumber
\end{eqnarray}
Using this equality and equation (\ref{eq40}) we obtain a recursion relation
for the total number of ASMs:
\begin{eqnarray}
&&A_n=\sum_{r=1}^n A(n,r)=\frac{A(n,1)}{(2n-2)!}\>\sum_{r=1}^n
\frac{(n+r-2)!\>(2n-r-1)!}{(r-1)!\>(n-r)!}.\nonumber
\end{eqnarray}
The sum on the right hand side of this equation can be found
by comparing the power series expansions on both sides of the next identity 
\footnote{
It is also possible to have a look at  standart textbooks on combinatorics.}
\begin{eqnarray}
&& x^{1-n}\>(1-x)^{-2n} \equiv (1-x)^{-n} \times x^{1-n}\>(1-x)^{-n}.\nonumber
\end{eqnarray}
As a result we get the famous recursion:
 \begin{eqnarray}
&&\frac{A_{n+1}}{A_n} = \frac{(3n+1)!\> n!}{(2n+1)!\> (2n)!}.\nonumber
\end{eqnarray}
Using $A_1=1$ we find the total number of $n \times n$ ASMs,
which was conjectured by Mills, Robbins, and Rumsey~\cite{MRR,MRR2}
and then was proved by Zeilberger~\cite{Z0} and Kuperberg~\cite{Ku1}.

\begin{center}{\bf5. Double  ASM distributions}
\end{center}

Consider now a more general case of two nonzero spectral parameters:
\begin{eqnarray}
\label{2case}
&&x_1=u\>\> \mbox{and} \>\>x_n=\tilde{u},  
\end{eqnarray}
which are are attached to the upper row and the lower row, respectively. 

Here the analogue of equation (\ref{1stZ}) for the state sum becomes 
\begin{eqnarray}
\label{eq42}
&&Z(n;u,\tilde{u})=\sum_{r,\tilde{r}=1}^{n}\>B(n;r,\tilde{r})\>a^{r-1}(u)\>b^{n-r}(u)\>
b^{\tilde{r}-1}(\tilde{u})\>a^{n-\tilde{r}}(\tilde{u}),
\end{eqnarray}
where $B(n,r,\tilde{r})$ is the number of ASMs of order $n$ whose sole '1' of the upper row,
is at the $r^{th}$ column and  whose sole '1' of the lower row, is at the $\tilde{r}^{th}$
column.

According to the symmetry, announced in Section 3, we have to get the same value of the state
sum by fixing the non-zero spectral parameters as follows: 
\begin{eqnarray}
&&x_1=u\>\> \mbox{and} \>\>y_1=\tilde{u}. \nonumber  
\end{eqnarray}
In this case the non-zero parameters are attached to the upper row and the left column
and we can present the state sum in a slightly different way
\begin{eqnarray}
\label{eq44}
&&Z(n;u,\tilde{u})= A(n;1)\>b^{n-1}(u)\>b^{n-1}(-\tilde{u})+a(u-\tilde{u})  \nonumber
\\
&&\sum_{r,\tilde{r}=2}^{n}
C(n;r,\tilde{r})\>a^{r-2}(u)\>b^{n-r}(u)\>a^{\tilde{r}-2}(-\tilde{u})\>
b^{n-\tilde{r}}(-\tilde{u}),
\end{eqnarray}
 where $C(n;r,\tilde{r}),\>\>r,\tilde{r} \ge 2$  is the number of ASMs  of order $n$ whose
 sole '1' in the upper row, is at the $r^{th}$ column and  whose sole '1' in the left column,
 is at the $\tilde{r}^{th}$ row.  $A(n,1)$ is the number of ASM with '1'
at the left and upper corner.

Equating (\ref{eq42}) and (\ref{eq44}), using equalities
$a(\pm
\tilde{u})=b(\mp \tilde{u})$ (see (\ref{upperBW})), and dividing the resulting equation by
$(b(u)\>a(\tilde{u}))^{n-1}$, we obtain
\begin{eqnarray}
\label{eq45}
&&\sum_{r,\tilde{r}=1}^{n}\>B(n;r,\tilde{r})\>\biggl(\frac{a(u)}{b(u)}\biggr)^{r-1}\>
\biggl(\frac{b(\tilde{u})}{a(\tilde{u})}\biggr)^{\tilde{r}-1} \equiv A(n,1)+ \nonumber
\\
&&+\frac{a(u-\tilde{u})}{b(u)\>a(\tilde{u})}\sum_{r,\tilde{r}=2}^{n}\>C(n;r,\tilde{r})
\>\biggl(\frac{a(u)}{b(u)}\biggr)^{r-2}
\biggl(\frac{b(\tilde{u})}{a(\tilde{u})}\biggr)^{\tilde{r}-2}.
\end{eqnarray}
Now we have to use the identity:
\begin{eqnarray}
&&\frac{a(u-\tilde{u})}{b(u)\>a(\tilde{u})}=\frac{a(u)}{b(u)}+\frac{b(\tilde{u})}
{a(\tilde{u})}-1. \nonumber
\end{eqnarray}
Writing $t$ and $\tilde{t}$ for the ratios $a(u)/b(u)$ and  $b(\tilde{u})/a(\tilde{u})$
respectively we can present equation (\ref{eq45}) in more transparent way
\begin{eqnarray}
&&\sum_{r,\tilde{r}=1}^{n}\>B(n;r,\tilde{r})\>t^{r-1}\>\tilde{t}^{\tilde{r}-1}\equiv
A(n,1)+ \nonumber \\
&&+(t+\tilde{t}-1)
\sum_{r,\tilde{r}=2}^{n}\>C(n;r,\tilde{r})\>t^{r-2}\>\tilde{t}^{\tilde{r}-2}.
\nonumber
\end{eqnarray}
Equating the coefficients we get
\begin{eqnarray}
&& A(n,1)=C(n;2,2), \quad r=\tilde{r}=1, \nonumber \\ 
&& B(n;r,\tilde{r})=C(n;r,\tilde{r}+1)+C(n;r+1,\tilde{r})-C(n;r+1,\tilde{r}+1), \nonumber \\
&&r+\tilde{r} > 2. \nonumber
\end{eqnarray}
The first equation  is a consequence of a simple bijection~\cite{Domino}.
The second equation for $\tilde r=1$ and $r>1$ gives the recursion relation for $C(n;r,2)$:
\begin{eqnarray} 
&& C(n;r,2)-C(n;r+1,2)= B(n;r,1)=A(n-1,r-1), \nonumber
\end{eqnarray}
where $A(n-1,r-1)$ is the number of ASM matrices $(n-1) \times (n-1)$ for which the
(unique) `1'
of the first row is at the $(r-1)^{th}$ column. This recursion relation can be obtained by a
simple bijection as well.
I reckon that the remaining relations ($r,\tilde r>1$) are new. 

In the final part of this section we intend to reduce the double ASM distribution
$B(n;r,\tilde r)$ to the numbers $A(n,r)$ (see equation (\ref{eq40}). 
In the case of two nonzero spectral parameters (see equation (\ref{2case})) we can rewrite
equation (\ref{eq14}) as:
\begin{eqnarray} 
&& f_{double}(u)=Z(n;u,\tilde u) \sin^{2n-2}u \sin(u-\tilde u).\nonumber
\end{eqnarray}
This function is the trigonometric polynomial of degree at most $3n-2$.
It can be found up to an arbitrary constant by using equations (\ref{eq15}) and
(\ref{addref}).
Let us consider  the combination 
\begin{eqnarray} 
&& f_{double}(u)=const (f(u) f^{\prime}(\tilde u)-f^{\prime}(u) f(\tilde u)), \nonumber
\end{eqnarray}
where the function $f(u)$ was found in Section 4.
This combination  satisfies all above-mentioned conditions and hence solves the problem. 
Inserting (\ref{fnew}) into the last equation we obtain after tedious  but straightforward
calculations:
\begin{eqnarray}
\label{eq51}
&&\sum_{r,\tilde{r}=1}^{n} B(n;r,\tilde{r}) t^{n-r}
\tilde{t}^{\tilde{r}-1}=\mbox{const}\>\frac{H(t)G(\tilde{t})-H(\tilde{t})G(t)}{t-\tilde
{t}}.
\end{eqnarray}  
We use two new functions:
\begin{eqnarray}
&&G(t)=(1-t)\>A(t), \nonumber \\
&&H(t)=(n-1/2)\>(1+t)\>A(t) +(1-n)(1-t)(1/2-t)\>A(t)-\nonumber \\
&&-(1-t)(t^2-t+1)\>A^{\prime}(t), \nonumber
\end{eqnarray}
where  $A(t)$ is the generating function defined by equation (\ref{eq38}).
Equation (\ref{eq51}) leads to a nonlinear relation between the `double refined' ASM
distributions and the numbers appearing in the refined ASM conjecture (theorem!):
\begin{eqnarray}
\label{last}
&& B(n;r+1,\tilde r+1)-B(n;r,\tilde r) =
\{A(n-1,r)[A(n,\tilde{r}+1)-A(n,\tilde{r})]+\nonumber \\
&&A(n-1,\tilde r)[A(n,r+1)-A(n,r)]\}/A(n,1).
\end{eqnarray}

{\it Acknowledgments} I would like to thank F.~C.~Alcaraz for his interest in this work
and for his valuable comments. I'm
grateful to N.~Slavnov for several useful discussions. 
The work was supported in part by the Brazilian agency FAPESP, by the Russian
Foundation for Basic Research under grant \# 01--01--00201 and by the INTAS under grant
\#
00--00561.

\it{Note added to the second ArXiv version of this paper}

\rm Combining equation (\ref{eq22}) with equation (\ref{eq2}) we can relate the
determinant $P$ with the
Izergin-Korepin determinant up to a constant multiplier.
The precise form of this relation can be derived \cite{Ok1}
by using the $\eta=2\pi/3$ specialization of the  theorem 4.2 of paper \cite{Ok2}.

\end{document}